\documentclass[aps,prl,twocolumn,superscriptaddress,longbibliography,floatfix]{revtex4-1}

\usepackage[T1]{fontenc}
\usepackage{newtxtext}
\usepackage{newtxmath}

\usepackage{amsmath}
\usepackage{mathtools}
\usepackage{amsfonts}
\usepackage{bm}

\usepackage{graphicx}
\usepackage{color}
\usepackage{xcolor}
\usepackage[colorlinks,
            citecolor=blue,
            linkcolor=red,
            urlcolor=blue]{hyperref}

\begin{document}

\title{Monolayer CrCl$_3$ as an ideal Test Bed for the Universality Classes of 2D Magnetism}

\author{M. Dupont}
    \affiliation{Department of Physics, University of California, Berkeley, California 94720, USA}
    \affiliation{Materials Sciences Division, Lawrence Berkeley National Laboratory, Berkeley, California 94720, USA}
\author{Y. O. Kvashnin}
    \affiliation{Department of Physics and Astronomy, Uppsala University, Box 516, S-751 20 Uppsala, Sweden}
\author{M. Shiranzaei}
    \affiliation{Department of Physics and Astronomy, Uppsala University, Box 516, S-751 20 Uppsala, Sweden}
\author{J. Fransson}
    \affiliation{Department of Physics and Astronomy, Uppsala University, Box 516, S-751 20 Uppsala, Sweden}
\author{N. Laflorencie}
    \affiliation{Laboratoire de Physique Th\'eorique, IRSAMC, Universit\'e de Toulouse, CNRS, UPS, 31062 Toulouse, France}
    \author{A.~Kantian}
    \affiliation{SUPA, Institute of Photonics and Quantum Sciences, Heriot-Watt University, Edinburgh, EH14 4AS, United Kingdom}
    \affiliation{Department of Physics and Astronomy, Uppsala University, Box 516, S-751 20 Uppsala, Sweden}

\begin{abstract}
    The monolayer halides CrX$_3$ ($X=$ Cl, Br, I) attract significant attention for realizing $2$D magnets with genuine long-range order (LRO), challenging the Mermin-Wagner theorem. Here, we show that monolayer CrCl$_3$ has the unique benefit of exhibiting tunable magnetic anisotropy upon applying a compressive strain. This opens the possibility to use CrCl$_3$ for producing and studying both ferromagnetic and antiferromagnetic $2$D Ising-type LRO as well as the Berezinskii-Kosterlitz-Thouless (BKT) regime of $2$D magnetism with quasi-LRO. Using state-of-the-art density functional theory, we explain how realistic compressive strain could be used to tune the monolayer's magnetic properties so that it could exhibit any of these phases. Building on large-scale quantum Monte Carlo simulations, we compute the phase diagram of strained CrCl$_3$, as well as the magnon spectrum with spin-wave theory. Our results highlight the eminent suitability of monolayer CrCl$_3$ to achieve very high BKT transition temperatures, around $50$ K, due to their singular dependence on the weak easy-plane anisotropy of the material.
\end{abstract}

\maketitle

\textit{Introduction.}--- Two-dimensional ($2$D) systems are of unique importance to many-body quantum mechanics, as attested, e.g., by superconductivity in the cuprates~\cite{keimer_quantum_2015} and at the LAO/STO interface~\cite{reyren_superconducting_2007}, as well as graphene monolayers~\cite{novoselov_two-dimensional_2005}. Part of this importance stems from the Mermin-Wagner (MW) theorem~\cite{PhysRevLett.17.1133,PhysRev.158.383}, which precludes any long-range order (LRO) arising from the spontaneous breaking of a continuous symmetry in two dimensions, but leaves room for a topological transition at finite temperature, named after Berezinskii~\cite{Berezinsky:1970fr}, Kosterlitz, and Thouless~\cite{kosterlitz_ordering_1973,Kosterlitz_1974} (BKT), e.g., in superfluid thin films~\cite{PhysRevLett.40.1727} or $2$D easy-plane (EP) magnets. Conversely, for easy-axis (EA) magnets, LRO due to the breaking of a discrete symmetry (e.g., $\mathbb{Z}_2$ for Ising systems) can occur at finite temperature. Such $2$D magnets are at the forefront of both experiment and theory, not only for these fundamental reasons, but also for applications, ranging from spintronics~\cite{Kapoor2020,Ningrum2020} to both classical~\cite{Song2018a} and quantum information~\cite{Ningrum2020}. Based on the precise demands, materials in different universality classes of magnetic order may be desired, each of which may face specific fundamental challenges to be realized in two dimensions. Recently, genuine magnetic LRO has been observed in chromium halides CrX$_3$ ($X=$ Cl, Br, I), which show local magnetic moments of spin greater than $1/2$, in the few --- and monolayer regimes~\cite{Gong2017, Huang2017}. These results had substantial impact and induced much follow-up work in many other atomically thin van der Waals materials and their heterostructures~\cite{Burch2018, Gibertini2019}.

First-principles density-functional theory (DFT) calculations show, and experiment confirms, that the insulating CrX$_3$ realizes highly localized magnetic moments close to $3\mu_\mathrm{B}$~\cite{PhysRevMaterials.1.014001, doi:10.1063/1.1714194}, corresponding to an ideal $S=3/2$ system, with short-range, Heisenberg-like superexchange couplings, as well as local EA magnetic anisotropy. The latter allows CrX$_3$ to overcome the MW theorem and to establish $2$D magnetic LRO~\cite{PhysRev.115.2,KANAMORI195987}. These traits imbue CrX$_3$ with a major advantage compared to gapless itinerant magnets~\cite{Fumega2019}. It is then CrCl$_3$ specifically that has unique potential for realizing magnetic universality classes beyond those with LRO, as it shows only a small EA anisotropy due to its lighter ligand, making it the most amenable to sign change by external manipulation, and thus realizing an EP anisotropy instead~\cite{PhysRevMaterials.1.014001}. DFT study further predicts that the anisotropy of the exchange in CrCl$_3$ is sufficiently suppressed~\cite{PhysRevB.102.115162}. In contrast, bulk and monolayers of CrBr$_3$ and CrI$_3$ are predicted to show strong EA anisotropy (both single-ion and intersite) for which achieving sign change would be unrealistic. This strong anisotropy arises from the spin-orbit coupling emerging from the heavier ligands~\cite{Lado_2017,Tartagliaeabb9379}, and CrI$_3$ further also displays strongly anisotropic exchange (possibly stemming from Kitaev interactions)~\cite{xu_interplay_2018,PhysRevLett.124.017201}.

Thus the opportunity to turn EA into EP anisotropy in CrCl$_3$ via compressive strain raises the possibility of tuning a material across strikingly different universality classes, with remarkable critical properties. Of greatest interest in this respect is the BKT regime, marked by the appearance of topological vortex excitations. Below a critical temperature the BKT regime exhibits quasi-LRO, i.e., with critical algebraic correlations. But while the realization of the quasi-LRO regime was first proposed for magnetic systems~\cite{kosterlitz_ordering_1973}, it has been surprisingly difficult to detect in such. In the various layered bulk magnets in which it is sought at low temperatures, there is invariably a temperature scale below which the weak coupling between the $2$D layers gives rise to an effective $3$D regime~\cite{PhysRevB.94.144403,PhysRevLett.124.177205} with its attendant magnetic LRO, obscuring the sought-after BKT physics~\cite{Tutsch2014,Hu2020a}. 
 
\begin{figure}[!t]
    \includegraphics[width=1\columnwidth]{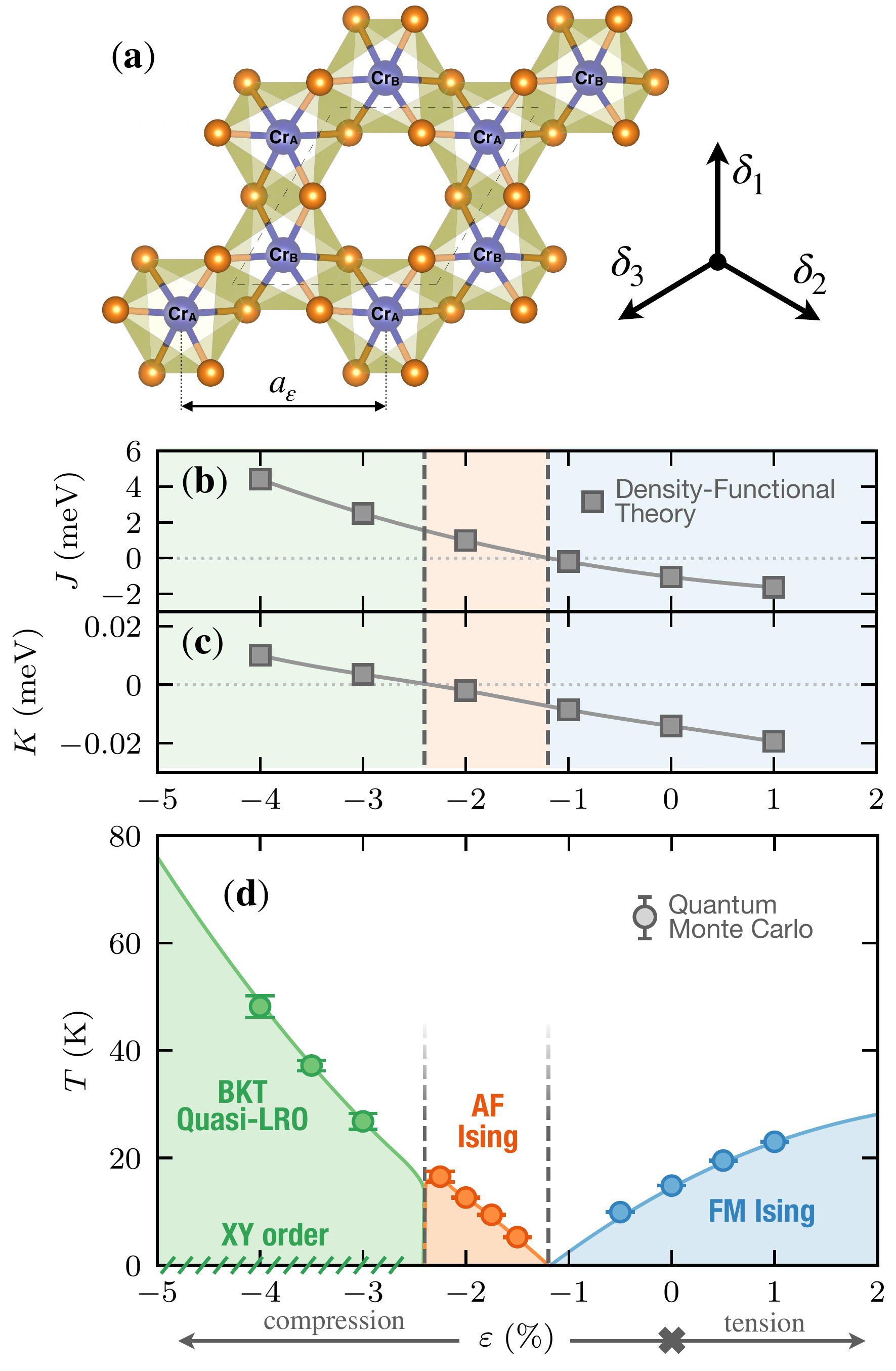} 
    \caption{\textbf{(a)} Crystal structure of monolayer CrCl$_3$. The Cr and Cl atoms are represented in blue and orange, respectively. The $\mathcal{A}$ and $\mathcal{B}$ sublattices of Cr is indicated. Dashed lines denote the unit cell with basic vectors $\boldsymbol{\delta}_1=a_\varepsilon\bigl(0, 1\bigr)$, $\boldsymbol{\delta}_2=a_\varepsilon\bigl(\sqrt{3}/2,-1/2\bigr)$, $\boldsymbol{\delta}_3=a_\varepsilon\bigl(-\sqrt{3}/2,-1/2\bigr)$, with strain-dependent lattice constant $a_\varepsilon$. \textbf{(b-c)} Magnetic nearest-neighbor superexchange $J$ and anisotropy $K$ of Hamiltonian~\eqref{eq:hamiltonian} respectively, computed via DFT as function of monolayer strain $\varepsilon$. \textbf{(d)} Finite-temperature phase diagram of the monolayer CrCl$_3$ versus strain $\varepsilon$, obtained by QMC simulations for the $S=3/2$ model of Eq.~\eqref{eq:hamiltonian} on the $2$D honeycomb lattice. Strain drives the monolayer into three different finite-temperature magnetic phases: BKT quasi-LRO phase for $\varepsilon\lesssim -2.4\%$, AFM Ising for $-2.4\%\lesssim\varepsilon\lesssim -1.2\%$, and FM Ising for $\varepsilon\gtrsim -1.2\%$. At zero temperature, the BKT quasi-LRO turns into genuine $XY$ LRO, separated from the AFM Ising order by an isotropic Heisenberg point displaying N\'eel order (where $K_\varepsilon$ vanishes). The AFM and FM Ising phases are separated by a trivial paramagnetic point (where $J_\varepsilon$ vanishes). The colored lines are fits to the form of Eq.~\eqref{eq:Tc}.}
    \label{fig:overview}
\end{figure}

In this Letter, we propose that, among the monolayer halides, CrCl$_3$ provides unique advantages for tuning material properties using, e.g., externally applied pressure such that both $2$D Ising ferromagnetic (FM) and antiferromagnetic (AFM) states with LRO as well as the sought-after BKT universality class could be observed, all in the same material. This ability is based on monolayer CrCl$_3$ realizing a $2$D spin-$3/2$ Heisenberg-like Hamiltonian with local anisotropy on a honeycomb lattice with high fidelity, where both nearest-neighbor superexchange coupling and magnetic anisotropy are susceptible to tuning of both their magnitude and sign due to strain $\varepsilon$ caused by realistic external pressure. Most pertinently, this in turn yields a BKT transition that is predicted to occur at much higher temperatures than true LRO in the layered bulk material~\cite{cable_neutron_1961}, found to be around $17$\;K at zero strain~\cite{doi:10.1021/acs.nanolett.0c02381}, see Fig.~\ref{fig:overview}\,(d). We further calculate the spin-excitation spectra of the material in the various strain regimes using the spin-wave approximation.

\textit{DFT of the monolayer and the effect of strain.---} The general crystal structure of monolayer CrCl$_3$ is depicted in Fig.~\ref{fig:overview}\,(a). Within our DFT approach, we first obtain the equilibrium structure of the crystal. Then, we strain the lattice while allowing Cl atoms to adjust their positions in order to minimize the energy cost of the lattice deformation at each chosen value of strain $\varepsilon$. For every structure generated, we compute the total energy difference between FM and AFM states. This allows us to extract the effective value of the nearest-neighbour exchange coupling $J_\varepsilon$ by mapping the energy difference onto Hamiltonian:
\begin{equation}
    \hat{H} = J_\varepsilon\sum_{\langle\boldsymbol{r},\boldsymbol{r'}\rangle} \boldsymbol{\hat{S}}_{\boldsymbol{r}} \cdot \boldsymbol{\hat{S}}_{\boldsymbol{r'}} + K_\varepsilon\sum_{\boldsymbol{r}}\left(\hat{S}^z_{\boldsymbol{r}}\right)^2,
    \label{eq:hamiltonian}
\end{equation}
where $\boldsymbol{\hat{S}}_{\boldsymbol{r}}=\bigl(\hat{S}^x_{\boldsymbol{r}},\hat{S}^y_{\boldsymbol{r}},\hat{S}^z_{\boldsymbol{r}}\bigr)$ are the standard $S=3/2$ spin-operators positioned on the vertices of a $2$D honeycomb lattice. The sum $\langle\boldsymbol{r},\boldsymbol{r'}\rangle$ restricts the magnetic exchange to nearest-neighbor spins. We perform fully relativistic calculations in order to compute the magnetic anisotropy $K_\varepsilon$, calculating the total energy difference between in-plane and out-of-plane orientations of the magnetization; for details, see the Supplemental Material (SM)~\cite{supplemental}. Our approach is in line with previous work studying the magnetic properties of Cr$X_3$~\cite{C5CP04835D,PhysRevB.100.205409,Lado_2017, PhysRevB.98.144411,PhysRevB.101.060404,xu_interplay_2018,olsen_2019,Pizzochero_2020}, which we note to yield critical temperatures in excellent agreement with experiment (see, e.g., Ref.~\onlinecite{doi:10.1021/acs.nanolett.0c02381}). Our own DFT treatment results in a nearest-neighbour distance for the Cr atoms of $3.424$\;{\AA} at $\varepsilon=0\%$, thus matching the measured bulk value of $3.44$\;{\AA}~\cite{crcl3-struct,cryst7050121}, signifying the accuracy of our approach.

We present our results for the Hamiltonian parameters $J_\varepsilon$ and $K_\varepsilon$ as a function of material strain $\varepsilon$ in Fig.~\ref{fig:overview}\,(b,\,c). It shows the FM configuration to be energetically favored at zero strain, in line with bulk CrCl$_3$, and the magnetic anisotropy to be of EA type and pointing out of plane, opposite to what is known for the bulk~\cite{PhysRevMaterials.1.014001}. This change in the monolayer limit has been obtained in prior DFT-based studies however~\cite{C5TC02840J, PhysRevB.98.144411,PhysRevB.100.224429}\footnote{Ref.~\onlinecite{PhysRevB.100.224429} suggests that the magnetic shape anisotropy in the unstrained monolayer CrCl$_3$ could overcome the magnetocrystalline anisotropy. We have not considered this effect, which would be problematic to treat in QMC.}. As compressive strain is applied to the monolayer, two key features of Fig.~\ref{fig:overview}\,(b,\,c) stand out: the sign change of $J_\varepsilon$ at $\varepsilon=-1.2\%$ from FM at AFM coupling, and of $K_\varepsilon$ at $\varepsilon=-2.4\%$ from EA to EP anisotropy as strain increases. These results validate our initial hypothesis that the much weaker magnetic anisotropy of monolayer CrCl$_3$ compared to CrI$_3$ and CrBr$_3$ offers an ideal platform to modify the Hamiltonian symmetry and thus explore Ising-type $2$D magnetism of both the FM and AFM variant (for $K_\varepsilon<0$), as well as the BKT regime (for $K_\varepsilon>0$), as the strains necessary are readily available in the lab; a strain of, e.g., $-4\%$ corresponds to pressure of $0.7$\;GPa. 

While a different choice of DFT exchange-correlation functional predicts a different $K_\varepsilon$~\cite{PhysRevB.98.144411}, we note that our own choice, also used in, e.g., Ref.~\onlinecite{PhysRevB.102.115162}, yields a better match to the experimentally found lattice constant, Cr-Cl distance and Cr-Cl-Cr bond angle at $\varepsilon = 0\%$, as well as yielding qualitatively the same phase diagram~\cite{supplemental}.

Our analysis reveals the source of the sign change in $J_\varepsilon$ as a subtle shift in balance between competing FM and AFM contributions, which we explicitly show in the Supplemental Material~\cite{supplemental}. According to the theory of superexchange~\cite{PhysRev.115.2,KANAMORI195987,GOODENOUGH1958287}, there is a FM superexchange between half-filled $t_{2g}$ and nominally empty $e_g$ orbital on the neighboring Cr atoms, mediated by a single Cl-$3p$ orbital~\cite{PhysRevB.99.104432,Kashin_2020,PhysRevB.102.115162}. This contribution is opposed by AFM superexchange between two different $t_{2g}$ orbitals via Cl state and also by the direct kinetic AFM exchange between the $t_{2g}$ orbitals pointing towards each other. Compressive strain on the monolayer decreases the Cr-Cr distance and increases the orbital overlap. While it is hard to say how superexchange paths are affected, the latter is definitely expected to boost the AF kinetic exchange term, which we argue to be the main driving force for the change of sign of $J_\varepsilon$ upon compressive strain.

\textit{From Ising to BKT.---} Building on the DFT-calculated couplings $J_\varepsilon$ and $K_\varepsilon$, we perform large-scale QMC simulations of the $S=3/2$ Hamiltonian of Eq.~\eqref{eq:hamiltonian}~\cite{supplemental}. We simulate $2$D systems of $N=2\times L\times L$ spins on the honeycomb lattice, up to $N\approx 5\times 10^4$, and map the phase diagram as a function of the strain $\varepsilon$ for CrCl$_3$, as shown in Fig.~\ref{fig:overview}\,(d).

\begin{figure}[t!]
    \center
    \includegraphics[width=1.0\columnwidth]{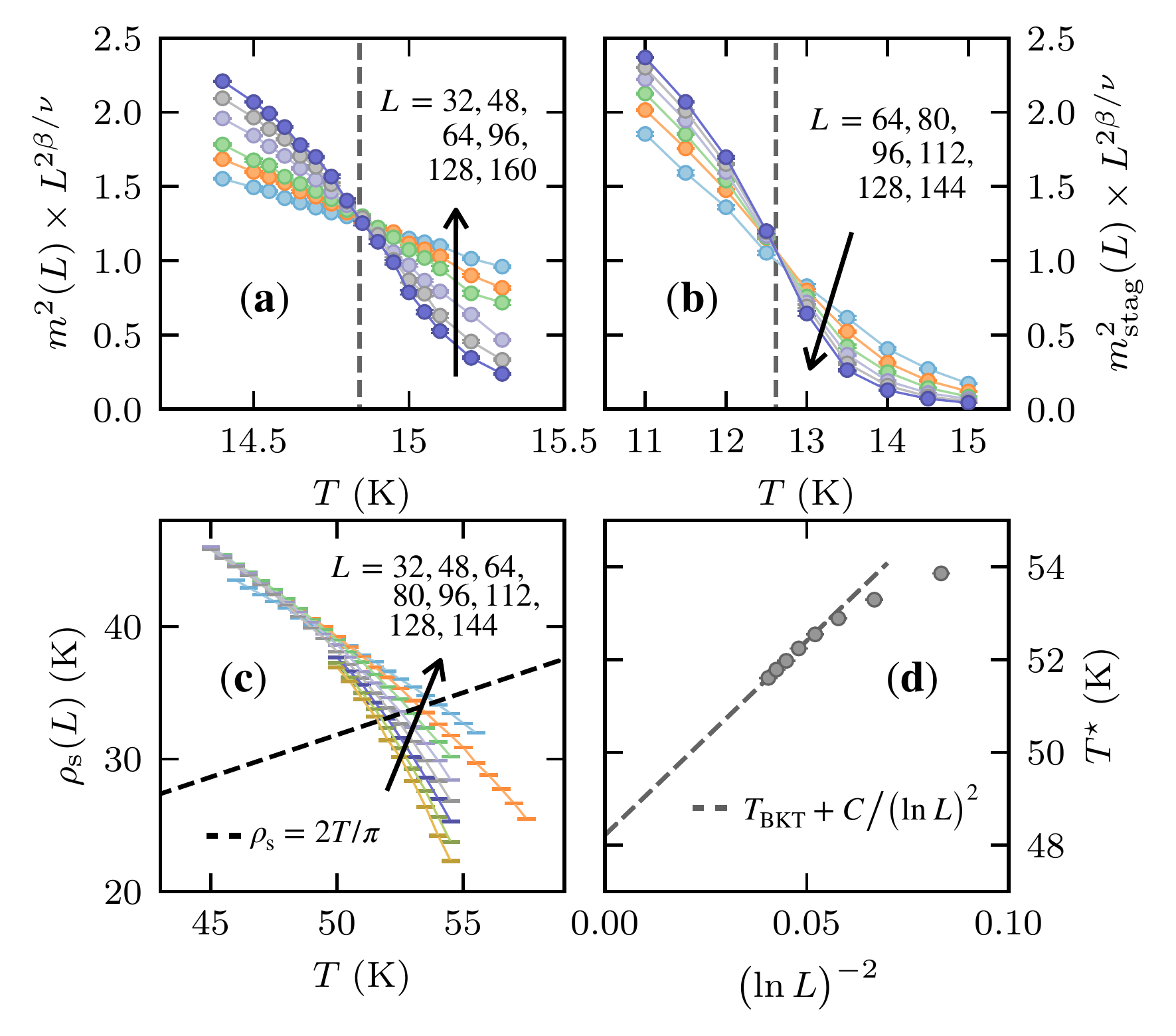} 
    \caption{\textbf{(a-b)} Obtaining $T_\mathrm{c}$ (dashed vertical lines) for onset of magnetic LRO from scaling analysis of QMC values of the order parameter in the EA regime, using $2$D Ising critical exponents $\beta=1/8$ and $\nu=1$. \textbf{(a)} Magnetization density vs. temperature for different $L$ at $\varepsilon=0\%$ (FM Ising), with $T_\mathrm{c}=14.84(1)\;\mathrm{K}$. \textbf{(b)} Staggered magnetization density vs. temperature for different $L$ at $\varepsilon=-2\%$ (AFM Ising), with $T_\mathrm{c}=12.6(1)\;\mathrm{K}$. \textbf{(c)} Finite-size scaling analysis of QMC-computed spin stiffness $\rho_\mathrm{s}(L)$ at different $L$ in EA regime, at $\varepsilon=-4\%$ (BKT Quasi-LRO). Dashed line shows $2T/\pi$. \textbf{(d)} $T^\star(L)$ extracted from (c) vs. $(\ln L)^{-2}$ for $\varepsilon=-4\%$. $T_\mathrm{BKT}=48(2)\;\mathrm{K}$ is extracted from fitting with $T_\mathrm{BKT}+C/(\ln L)^2$ (dashed line).}
    \label{fig:universality}
\end{figure}

For $K_\varepsilon<0$ ($\varepsilon>-2.4\%$) in the EA regime, we perform a finite-size scaling analysis of the magnetic order parameter in order to extract the critical temperature $T_\mathrm{c}$ for the onset of magnetic LRO, perfectly supporting the $2$D Ising universality class. This is exemplified in  Figs.~\ref{fig:universality}\,(a,b) where we find the thermal melting of both FM order for $\varepsilon>-1.2\%$, and AFM order for $\varepsilon\in[-2.4\%,-1.2\%]$, to be precisely described by the critical exponents $\beta=1/8$ for the order parameter, and $\nu=1$ for the correlation length~\cite{cardy1996scaling}, allowing for accurate extraction of $T_\mathrm{c}$.

When entering the EP regime for $K_\varepsilon>0$ ($\varepsilon<-2.4\%$), there is a drastic change in the critical properties. At zero temperature, true LRO is expected, breaking the $\mathrm{U}(1)$ symmetry, but at finite temperature the MW theorem precludes this~\cite{PhysRevLett.17.1133,PhysRev.158.383}, allowing at most for quasi-LRO in the $XY$ plane. This is what we observe for $\varepsilon=-4\%$ in Fig.~\ref{fig:universality}\,(c), where the system displays a finite spin stiffness $\rho_\mathrm{s}(T)$ below a transition temperature $T_\mathrm{BKT}\sim 50$\;K. Another manifestation of the transition to quasi-LRO is the onset of algebraic decay of spin correlations~\cite{supplemental}. We determine $T_\mathrm{BKT}$ both from the universal relation ${\rho_\mathrm{s}(T=T_\mathrm{BKT})=2T_\mathrm{BKT}/\pi}$~\cite{PhysRevLett.39.1201}, see Fig.~\ref{fig:universality}\,(c), as well as from critical correlations decaying with a universal exponent $\eta=1/4$~\cite{Kosterlitz_1974,supplemental}. Yet, strong logarithmic finite-size corrections are expected for BKT transitions~\cite{Kosterlitz_1974,2001EPJB,Hsieh_2013}, calling for a careful analysis. Noting $T^\star(L)$ the solution of $\rho_\mathrm{s}(L)=2\pi/T$, we extract the thermodynamic limit estimate of $T_\mathrm{BKT}$ through the relation~\cite{Bramwell_1993} $T^\star(L)=T_\mathrm{BKT}+C/(\ln L)^2$, valid as $L\to+\infty$, with $C$ a nonuniversal constant, as exemplified in Fig.~\ref{fig:universality}\,(d) for $\varepsilon=-4\%$, where we obtain $T_\mathrm{BKT}=48(2)$\;K. Several estimates are reported in Fig.~\ref{fig:overview}\,(d), where we observe a strong enhancement of $T_\mathrm{BKT}$ upon compressive strain for $\varepsilon<-2.4\%$. This remarkable increase is not directly controlled by the $\mathrm{SU}(2)\to \mathrm{U}(1)$ symmetry breaking term $K_\varepsilon$ in the Hamiltonian Eq.~\eqref{eq:hamiltonian}, but emerges from a strong non-linear effect, as we discuss now.

\begin{figure}
    \includegraphics[width=\columnwidth]{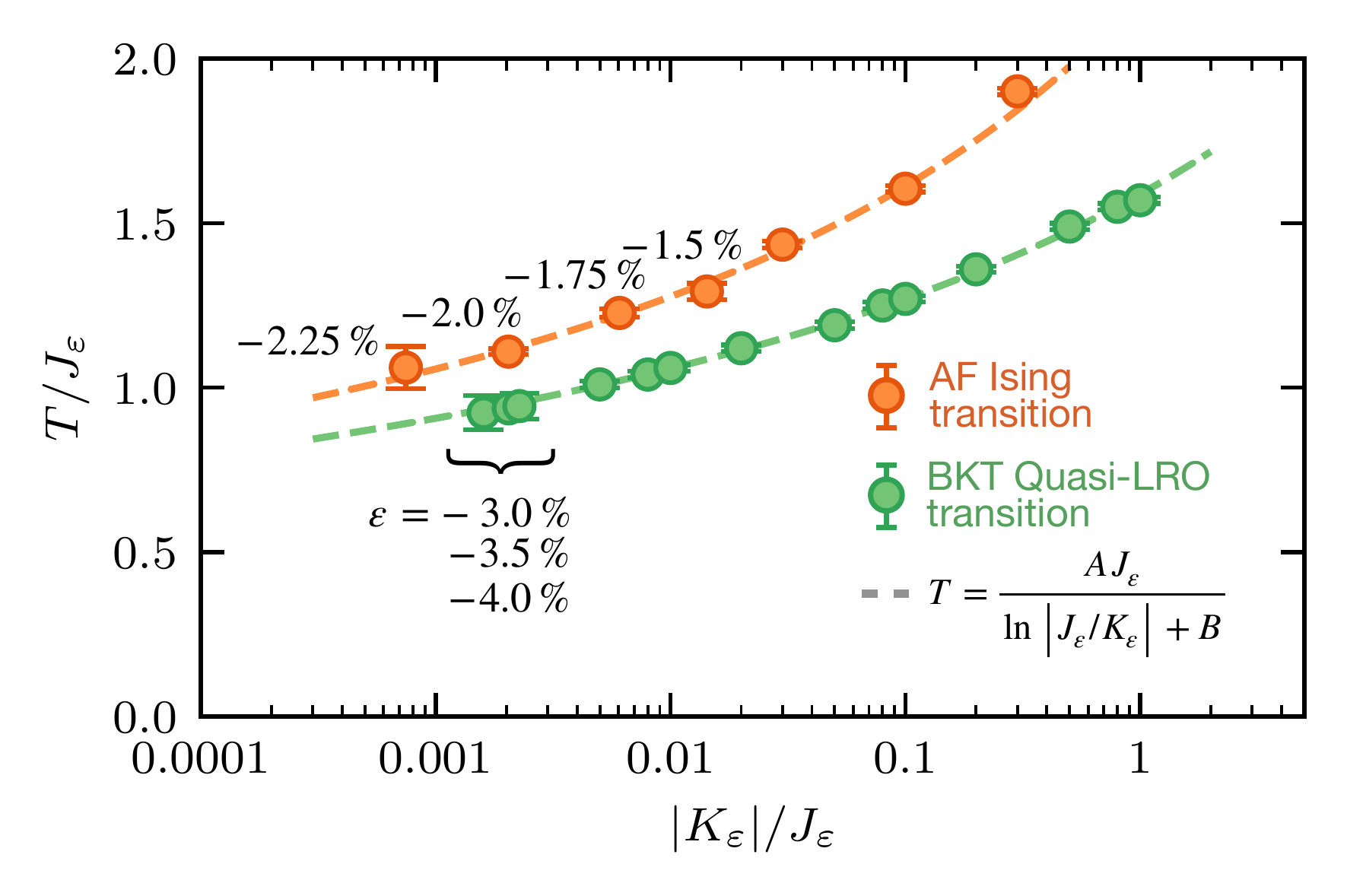} 
    \caption{Critical temperature $T_\mathrm{c}$ (AFM Ising, orange) and $T_\mathrm{BKT}$ (BKT Quasi-LRO, green) vs. $|K_\varepsilon|/J_\varepsilon$. Points with explicit $\varepsilon$ values correspond to $K_\varepsilon/J_\varepsilon$ for monolayer CrCl$_3$ at such strain, see Fig.~\ref{fig:overview}\,(d). Dashed lines show fit to $A/\bigl(\ln|J_\varepsilon/K_\varepsilon| + B\bigr)$ in the small anisotropy limit $|K_\varepsilon|/J_\varepsilon\leq 0.1$, with $A$ and $B$ fitting parameters. For BKT, $A=14.6(3)$, $B=9.2(3)$. For AFM Ising, $A=14.1(5)$, $B=6.5(4)$. Both $A$ values are compatible with the analytical prediction $A=4\pi\,\rho_\mathrm{s}^{(0)}=16(1)$ in Eq.~\eqref{eq:Tc}.}
    \label{fig:tc_scaling}
\end{figure}

\textit{Logarithmic enhancement of the critical temperature.---} The critical nature of $2$D systems at low temperatures results in a strong sensitivity to even weak anisotropies ($|K_\varepsilon/J_\varepsilon|$ is typically less than $\approx 10^{-2}$) that nudge the system towards a certain (quasi-)order. Thus, in line with previous work on alternative realizations of $2$D magnets ~\cite{Khokhlachev76,IRKHIN1997143,Roscilde2003}, we find a strong logarithmic enhancement of critical temperatures, which are controlled by the exchange $J_\varepsilon$ but also with a singular dependence on $K_\varepsilon$, both on the Ising and on the BKT side
for the CrCl$_3$ monolayer, as clearly shown in Fig.~\ref{fig:tc_scaling}. Using QMC for both physical parameters at various strains, as well as a broader range for the ratio $|K_\varepsilon|/J_\varepsilon$, we find excellent agreement with
\begin{equation}
    T_\mathrm{c,BKT}=\frac{4\pi J_\varepsilon\,\rho_\mathrm{s}^{(0)}}{\ln\left|{J_\varepsilon}\bigl/{K_{\varepsilon}}\right|+B},
    \label{eq:Tc}
\end{equation}
where $\rho_\mathrm{s}^{(0)}$ is the dimensionless spin stiffness of the isotropic ($K_\varepsilon=0$) system at zero temperature and $B$ is a nonuniversal constant~\cite{IRKHIN1997143,Roscilde2003}. A QMC estimate of the isotropic stiffness gives a prefactor $4\pi\rho_\mathrm{s}^{(0)}=16(1)$, which agrees well with our results displayed in Fig.~\ref{fig:tc_scaling}.

\textit{Magnon-spectra of the monolayer.---} Complementing our QMC description of the equilibrium properties we use spin-wave (SW) analysis for an $N$-site cluster of the system Hamiltonian~\eqref{eq:hamiltonian} to obtain predictions for the $T=0$ excitation spectrums of the monolayer at $\varepsilon=0\%$ (FM phase) and $\varepsilon=-4\%$ ($XY$ order), see Fig.~\ref{fig:overview}\,(d). In each case we model the deviations of the spins around a classical configuration, with the analytical procedure depending on whether this configuration is externally proscribed or has to be picked randomly. 

\begin{figure}
    \includegraphics[width=\columnwidth]{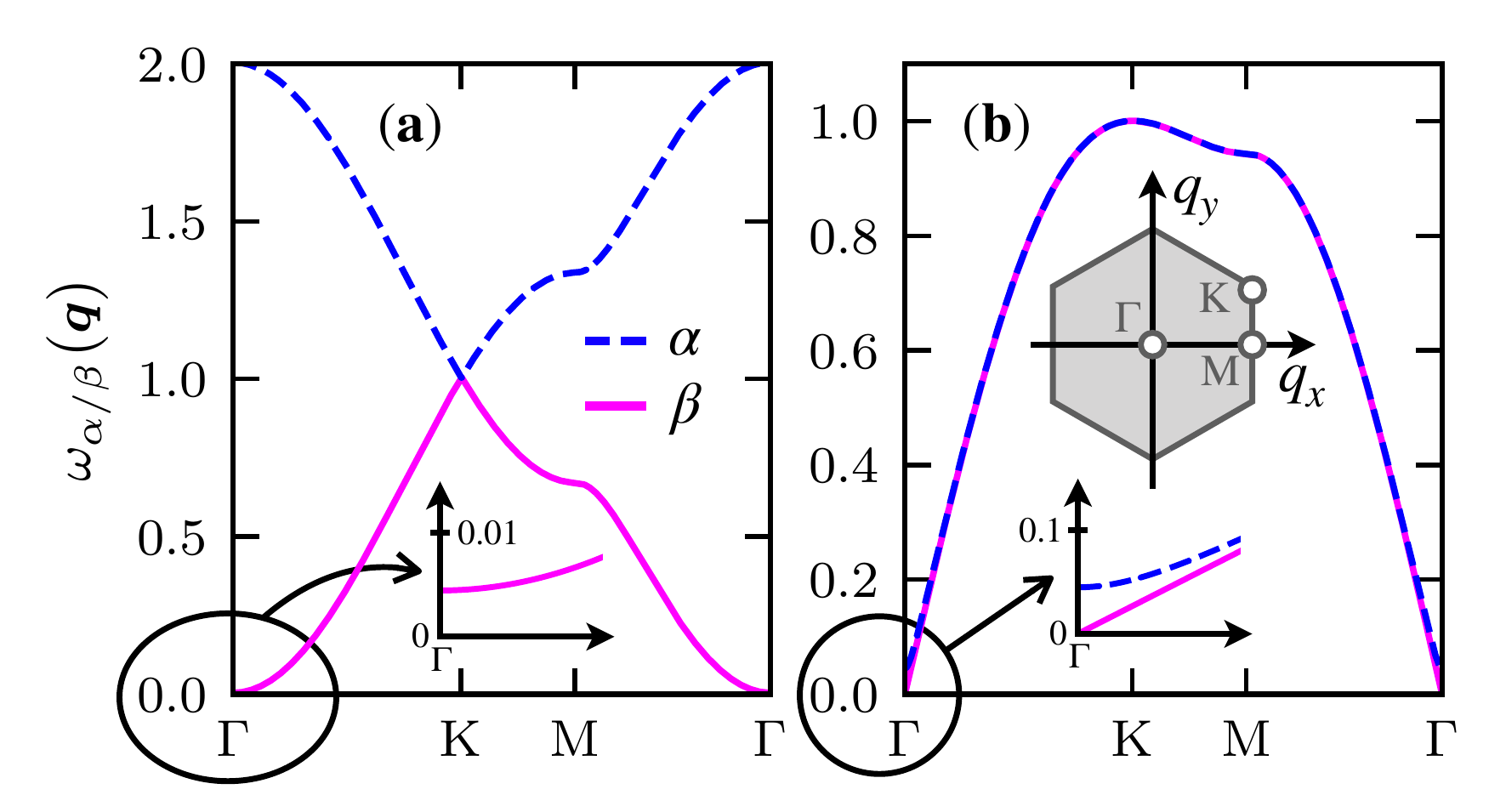} 
    \caption{Band structure of magnons in \textbf{(a)} easy-axis FM and \textbf{(b)} easy-plane AFM. The blue and magenta curves represent $\alpha$ and $\beta$ bands, respectively. Inset in panel \textbf{(a)} shows gap opening at the bottom of spectrum close to $\Gamma$ due to anisotropy while the one in panel \textbf{(b)} displays breaking of band degeneracy close to $\Gamma$. The Brillouin zone of the honeycomb lattice is displayed in \textbf{(b)}, along with the high symmetry points $\boldsymbol{q}=a_\varepsilon^{-1}\bigl(q_x,q_y\bigr)$: $\Gamma=a_\varepsilon^{-1}\bigl(0,0\bigr)$, $\mathrm{M}=a_\varepsilon^{-1}\bigl(4\pi/3,0\bigr)$, and $\mathrm{K}=a_\varepsilon^{-1}\bigl(\pi,\pi/\sqrt{3}\bigr)$ with strain-dependent lattice constant $a_\varepsilon$.}
    \label{fig:spectrum}
\end{figure}

For $K_\varepsilon<0$ ($\varepsilon=0\%$), the ground-state configuration is of the spins aligned out of plane in the $z$ direction. We use the appropriate Holstein-Primakoff (HP) transformation~\cite{PhysRev.58.1098} for mapping the magnon excitations above the ground state onto noninteracting bosons~\cite{supplemental}. Up to quadratic terms and after canonical transformations, one arrives at
\begin{align}
        \frac{\hat{H}}{NS^2}\approx& -\frac{3J_\varepsilon}{2} - K_\varepsilon + \frac{3J_\varepsilon}{NS} \sum_{\boldsymbol{q}}\Bigl[\omega_\alpha\bigl(\boldsymbol{q}\bigr) \hat{\alpha}^\dag_{\boldsymbol{q}} \hat{\alpha}_{\boldsymbol{q}}^{\vphantom{\dagger}}
        + \omega_\beta\bigl(\boldsymbol{q}\bigr)\hat{\beta}^\dag_{\boldsymbol{q}}  \hat{\beta}_{\boldsymbol{q}}^{\vphantom{\dagger}}\Bigr],
    \label{FMHamiltonian}
\end{align}
for approximating the magnon spectrum as the dispersion of two distinct types of free bosons, with 
\begin{equation}
    \omega_{\alpha/\beta}\bigl(\boldsymbol{q}\bigr) = 1 + \frac{K_\varepsilon}{3J_\varepsilon} \pm \left|\gamma\bigl(\boldsymbol{q}\bigr)\right|,
    \label{FMdispersion}
\end{equation}
where $\gamma(\boldsymbol{q})=(1/3)\sum_{n=1}^3\mathrm{e}^{i\boldsymbol{q}\cdot\boldsymbol{\delta}_n}$ and $\boldsymbol{\delta}_n$-vectors as shown in Fig.~\ref{fig:overview}\,(a). The EA anisotropy is seen to open a gap at the bottom of the lower $\beta$ branch, which stabilizes the system against the long-wavelength Goldstone modes that would otherwise result in the destruction of magnetic LRO, see Fig~\ref{fig:spectrum}\,(a).

For $K_\varepsilon>0$ ($\varepsilon=-4\%$), when the anisotropy becomes EP due to compressive strain, one has to pick an arbitrary orientation in the $XY$ plane along which the spins order; we chose the $x$ direction in the following. For the concrete CrCl$_3$ monolayer this procedure is justified by our DFT results, which show energy differences for different in-plane orientations to be well below the $\mu$eV level. Application of the standard HP-approach would violate the Goldstone theorem, so we use the matching of matrix-elements (MME) technique instead~\cite{supplemental, Lindgard_1976, Balucani_1980}. As spin-exchange dominates, we expand to the first power of $d_\varepsilon=K_\varepsilon/6J_\varepsilon S$. This results in another magnon-Hamiltonian structurally analogous to the one presented in Eq.~\eqref{FMHamiltonian}, but with the ground state energy replaced by ${NS\big[-3J_\varepsilon S+K_\varepsilon\big(1-d_\varepsilon(2S-1)\big)\big]/2}$ and the dispersion relation 
\begin{equation}
    \omega_{\alpha/\beta}\bigl(\boldsymbol{q}\bigr) = \sqrt{\Bigl[1+d_\varepsilon(2S-1)\left(1\pm\bigl| \gamma\bigl(\boldsymbol{q}\bigr)\bigr|\right)\Bigr]^2-\bigl|\gamma\bigl(\boldsymbol{q}\bigr)\bigr|^2}.
    \label{AFMdispersion}
\end{equation}
As shown in Fig.~\ref{fig:spectrum}\,(b), the magnon spectrum now is mostly degenerate except around the $\Gamma$ point where the EP anisotropy breaks the degeneracy, gapping the $\alpha$ branch while the $\beta$ branch remains linear down to $\boldsymbol{q}=\boldsymbol{0}$, thus signaling the presence of a Nambu-Goldstone mode associated with the breaking of the $\mathrm{U}(1)$ symmetry.

\textit{Discussion and outlook.---} Various signatures would be available in order to detect the transition of monolayer CrCl$_3$ to the BKT regime that we have shown to happen as compressive strain is increased. The decay-behavior of an induced spin-current as, e.g., emanating from a Pt electrode has been proposed for this~\cite{Kim2020}. So has been the minimum in the uniform magnetic $zz$ susceptibility predicted to occur for $2$D EP magnets just above the BKT transition, as opposed to the monotonous decline predicted for the $xx$ susceptibility~\cite{Cuccoli2003}. In that respect, the $2$D honeycomb $S=1$ AFM compound BaNi$_2$V$_2$O$_8$~\cite{PhysRevB.65.144443,PhysRevLett.91.137601} was very recently shown to be a good candidate~\cite{klyushina2020signatures}. These susceptibilities are practically accessible in current experiments on compressed monolayers. Crucially, the monolayer of CrCl$_3$ would not suffer from some intervening onset of $3$D magnetism which has obscured experimental studies of the BKT regime in layered bulk magnets thus far. The sum of the present work shows this material to be ideally situated in parameter space in order to address the major universality classes of $2$D magnetism with great control and accuracy.

\begin{acknowledgments}
    A. K. would like to thank M. Abdel-Hafiez for fruitful discussions. M. D. was supported by the U.S. Department of Energy, Office of Science, Office of Basic Energy Sciences, Materials Sciences and Engineering Division under Contract No. DE-AC02-05-CH11231 through the Scientific Discovery through Advanced Computing (SciDAC) program (KC23DAC Topological and Correlated Matter via Tensor Networks and Quantum Monte Carlo). Y. O. K. (project No. 2019-03569) and J. F. acknowledge financial support from Swedish Research Council (VR). M. Sh. and J. F. thank Carl Tryggers Stiftelse for financial support. N. L. acknowledges the French National Research Agency (ANR) under Projects THERMOLOC ANR-16-CE30-0023-02, and GLADYS ANR-19-CE30-0013. This research used the Lawrencium computational cluster resource provided by the IT Division at the Lawrence Berkeley National Laboratory (Supported by the Director, Office of Science, Office of Basic Energy Sciences, of the U.S. Department of Energy under Contract No. DE-AC02-05CH11231). This research also used resources of the National Energy Research Scientific Computing Center (NERSC), a U.S. Department of Energy Office of Science User Facility operated under Contract No. DE-AC02-05CH11231. The DFT computations were performed using the resources provided by the Swedish National Infrastructure for Computing (SNIC) at the National Supercomputing Centre (NSC).
\end{acknowledgments}

\bibliographystyle{apsrev4-1}
\bibliography{mybib}

\setcounter{secnumdepth}{3}
\setcounter{figure}{0}
\setcounter{equation}{0}
\renewcommand\thefigure{S\arabic{figure}}
\renewcommand\theequation{S\arabic{equation}}
\newpage\clearpage

\begin{center}
    \bfseries\Large Supplemental material to ``Monolayer CrCl$_3$ as an ideal Test Bed for the Universality Classes of 2D Magnetism''
\end{center}

{\noindent\itshape In this supplemental material, we provide technical information regarding the DFT calculations, the QMC simulations, and the spin-wave analysis. We also provide details regarding the finite-size scaling analyses of the QMC data for the FM Ising, AFM Ising, and BKT regimes. Finally, we provide a discussion on the different exchange paths in CrCl$_3$ and additional DFT results using the PBE functional.}

\section{Details of the DFT-calculations}
\label{sec:app:dft}

The equilibrium crystal structure of the monolayered CrCl$_3$ having $D_{3d}$ symmetry was obtained by performing a complete optimization within density functional theory (DFT)-based calculations. A $20\AA$-thick vacuum was added to ensure no interaction between the layers. 
The lattice parameters and atomic positions relaxed using projector augmented wave method as implemented in VASP code~\cite{paw,vasp}. We chose PBESol~\cite{PhysRevLett.100.136406} as exchange-correlation functional. The plane-wave kinetic energy cut-off for was set to $450$ eV along with $25\times 25\times1$ k-point grid. The forces on each atom were minimized down to $0.1~\mathrm{meV}/$\AA. Once the equilibrium structure was obtained, we have strained the lattice while allowing Cl atoms to adjust their positions in order to minimize the energy cost of the lattice deformation. This procedure was done for each chosen value of strain.

\section{Quantum Monte Carlo simulations}
\label{sec:app:qmc}

\subsection{Method and definitions}

Quantum Monte Carlo simulations of the $S=3/2$ lattice Hamiltonian of Eq.~(1) in the main text are performed with stochastic series expansion (SSE) algorithm using directed loops updates~\cite{PhysRevE.66.046701,PhysRevE.71.036706}. Our calculations are based on the ALPS library~\cite{Bauer2011a}. We simulate fully periodic $2$D systems of $N=2\times L\times L$ sites up to $\approx 5\cdot 10^4$ spins on the lattice. The phase boundaries of the model (strain versus temperature) are determined by a standard finite-size scaling analysis. We denote the Monte Carlo average by $\langle\cdot\rangle$. For the FM Ising transition (see Fig.~\ref{fig:nn_bond_4} of the main text), we consider the square of the magnetization density, defined as,
\begin{equation}
    m^2 = \left\langle\left(\frac{1}{N}\sum\nolimits_{\boldsymbol{r}}\hat{S}^z_{\boldsymbol{r}}\right)^2\right\rangle.
    \label{eq:fm_op}
\end{equation}
For the AFM Ising transition, we consider the square of the staggered magnetization density (see Fig.~\ref{fig:nn_bond_4} of the main text). It reads,
\begin{equation}
    m_\mathrm{stag}^2 = \left\langle\frac{1}{N^2}\left(\sum\nolimits_{\boldsymbol{r}\in\mathcal{A}}\hat{S}^z_{\boldsymbol{r}}-\sum\nolimits_{\boldsymbol{r}\in\mathcal{B}}\hat{S}^z_{\boldsymbol{r}}\right)^2\right\rangle,
    \label{eq:af_op}
\end{equation}
with $\mathcal{A}$ and $\mathcal{B}$ corresponding to the two sublattices of the hexagonal lattice, respectively. For the BKT transition, we consider two different quantities. First, the two-point off-diagonal spin correlation function at the longest distance $\boldsymbol{r_\mathrm{max}}$ available between two spins (i.e., $\|\boldsymbol{r_\mathrm{max}}\|\propto L$): $\bigl\langle\hat{S}^+_{\boldsymbol{r}}\hat{S}^-_{\boldsymbol{r}+\boldsymbol{r_\mathrm{max}}}\bigr\rangle$. We evaluate it while constructing the loop update~\cite{PhysRevE.64.066701} and average over all lattice sites $\boldsymbol{r}$. The second quantity we consider regarding the BKT transition is the spin stiffness $\rho_\mathrm{s}$, made readily accessible when expressed as the fluctuation of the winding number~\cite{PhysRevB.36.8343,PhysRevB.56.11678}.

\subsection{Ising transitions}

\begin{figure}[t]
    \includegraphics[width=\columnwidth]{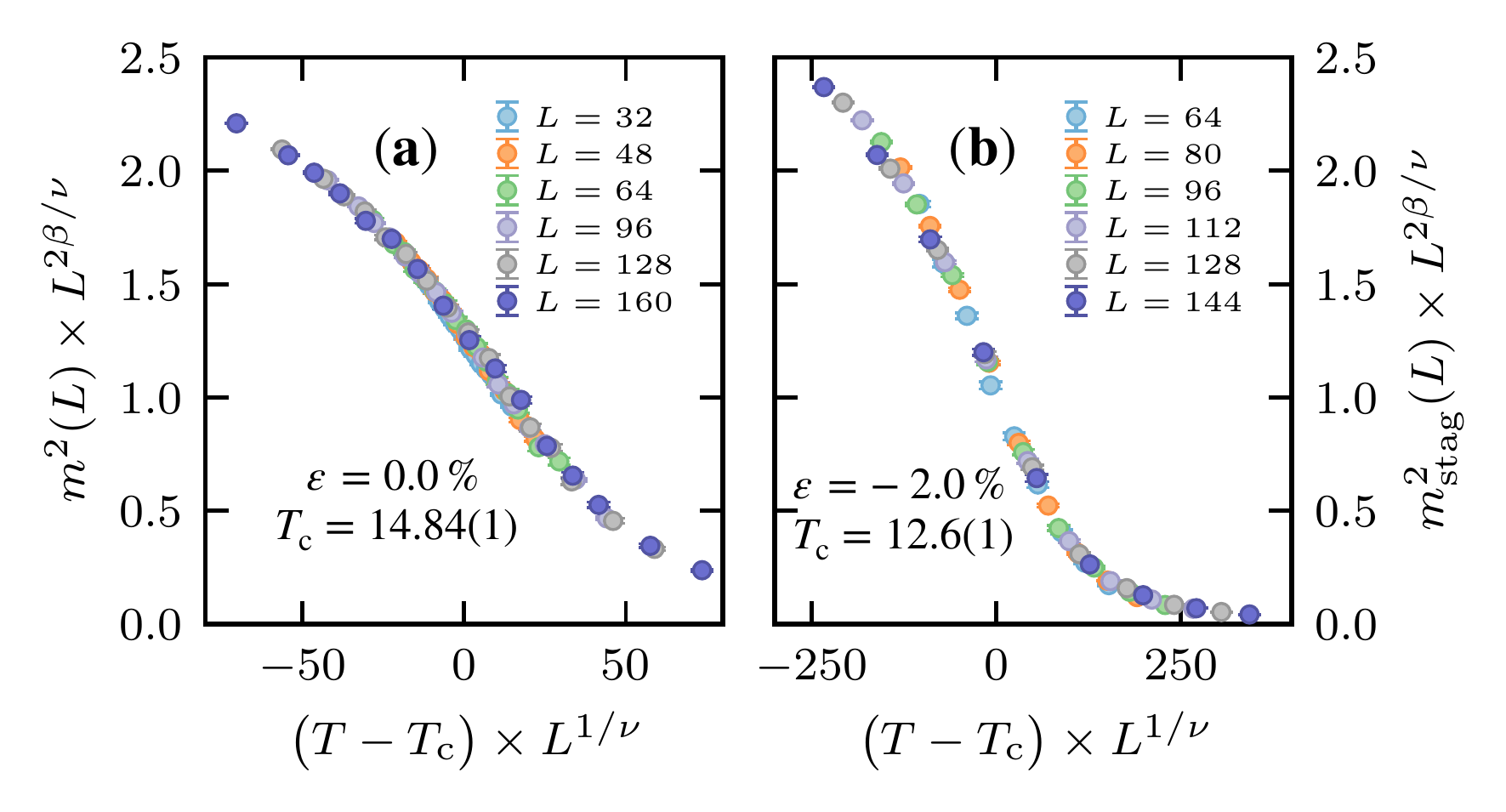} 
    \caption{Data collapse of Fig.~\ref{fig:nn_bond_4}\,(a,b) of the main text by rescaling the $x$ axis as $T\to(T-T_\mathrm{c})\times L^{1/\nu}$ with $\nu=1$ the correlation length critical exponent of the $2$D Ising universality class. \textbf{(a)} Magnetization density vs. temperature for different $L$ at $\varepsilon=0\%$ (FM Ising), with $T_\mathrm{c}=14.84(1)\;\mathrm{K}$. \textbf{(b)} Staggered magnetization density vs. temperature for different $L$ at $\varepsilon=-2\%$ (AFM Ising), with $T_\mathrm{c}=12.6(1)\;\mathrm{K}$.}
    \label{fig:ising_tc}
\end{figure}

To determine the critical temperature $T_\mathrm{c}$ for the FM and AFM Ising regimes, we use a standard finite-size scaling of the order parameter. For the FM Ising transition, the square of the magnetization density of Eq.~\eqref{eq:fm_op} follows, 
\begin{equation}
    m^2(L)=L^{-2\beta/\nu}\times\mathcal{F}_{m^2}\Bigl[\bigl(T-T\mathrm{c}\bigr)\times L^{1/\nu}\Bigr].
    \label{eq:scaling_fm}
\end{equation}
For the AFM Ising transition, the square of the staggered magnetization density of Eq.~\eqref{eq:af_op} follows, 
\begin{equation}
    m_\mathrm{stag}^2(L)=L^{-2\beta/\nu}\times\mathcal{F}_{m_\mathrm{stag}^2}\Bigl[\bigl(T-T\mathrm{c}\bigr)\times L^{1/\nu}\Bigr],
    \label{eq:scaling_af}
\end{equation}
with $\beta$ the order parameter and $\nu$ the correlation length exponents, respectively. For the $2$D Ising universality class, $\beta=1/8$ and $\nu=1$~\cite{cardy1996scaling}. $\mathcal{F}_{m^2}$ and $\mathcal{F}_{m_\mathrm{stag}^2}$ are universal scaling functions.

Throughout the paper, we determine the critical temperature of the AFM and FM regimes according to Eqs.~\eqref{eq:scaling_fm} and~\eqref{eq:scaling_af}. See Fig.~\ref{fig:nn_bond_4}\,(a,b) of the main text as well as Fig.~\ref{fig:ising_tc}, where the collapse of the data points correspond to the universal scaling functions. The perfect collapse using the $2$D Ising universality class exponents confirm the nature of the phase transition.

\subsection{Determining the BKT transition}

\begin{figure}[t]
    \includegraphics[width=\columnwidth]{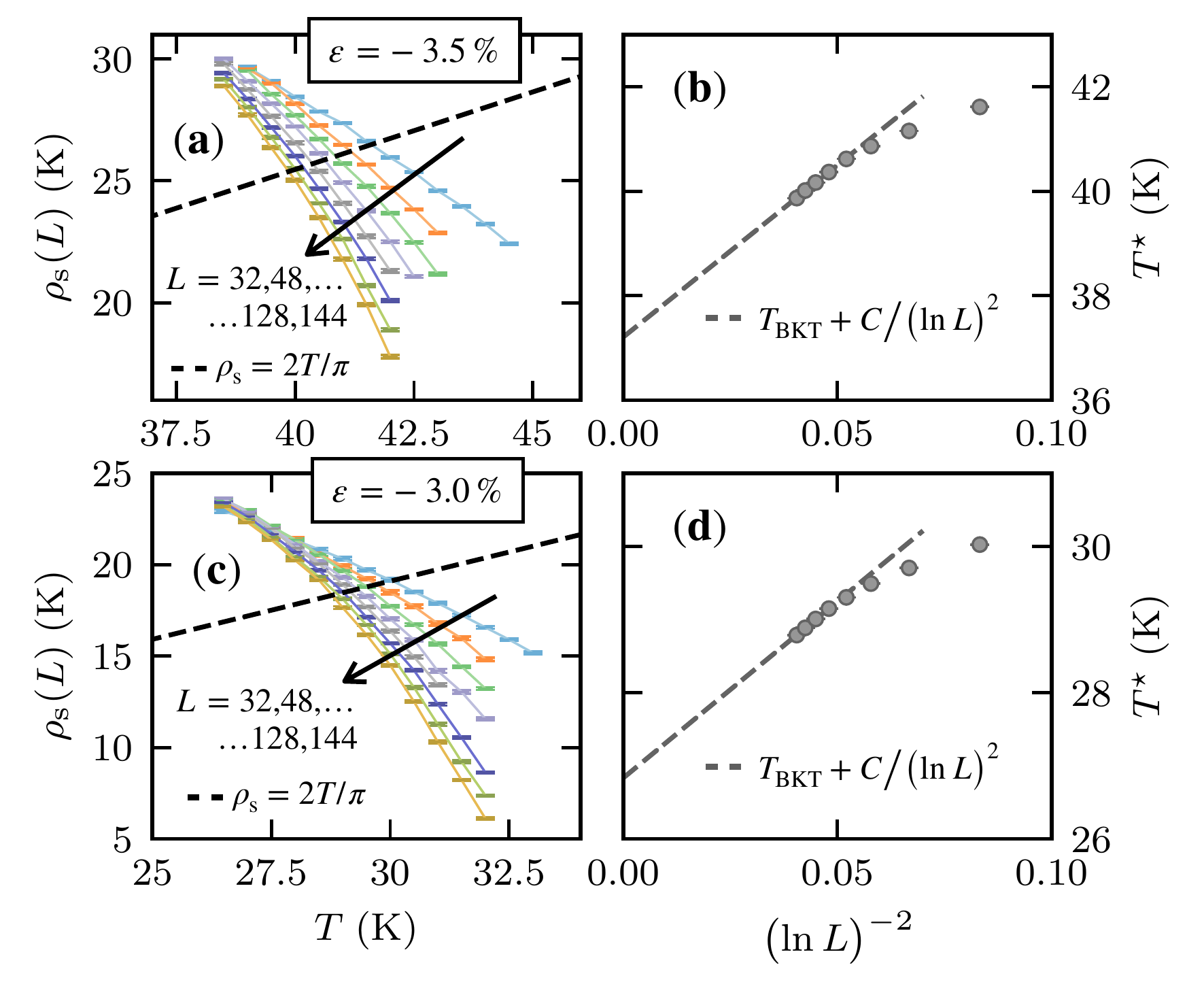} 
    \caption{Finite-size scaling analysis of the spin stiffness $\rho_\mathrm{s}(L)$ for different linear system sizes $L$ from QMC simulations. The hamiltonian parameters are those corresponding to $\varepsilon=-3.5\%$ (upper row) and $\varepsilon=-3.0\%$ (lower row). In the thermodynamic limit, at the BKT transition $T_\mathrm{BKT}$, the spin stiffness shows a universal jump and takes the value $\rho_\mathrm{s}=2\pi/T_\mathrm{BKT}$. For a finite-size system, we note $T^\star(L)$ the solution of $\rho_\mathrm{s}(L)=2\pi/T$. $T^\star(L)$ is expected to behave as $T_\mathrm{BKT}+C/(\ln L)^2$ for $L\to+\infty$, with $C$ a non-universal constant. We fit the data points to determine $T_\mathrm{BKT}$.}
    \label{fig:kt_stiffness}
\end{figure}

\begin{figure}[t]
    \includegraphics[width=\columnwidth]{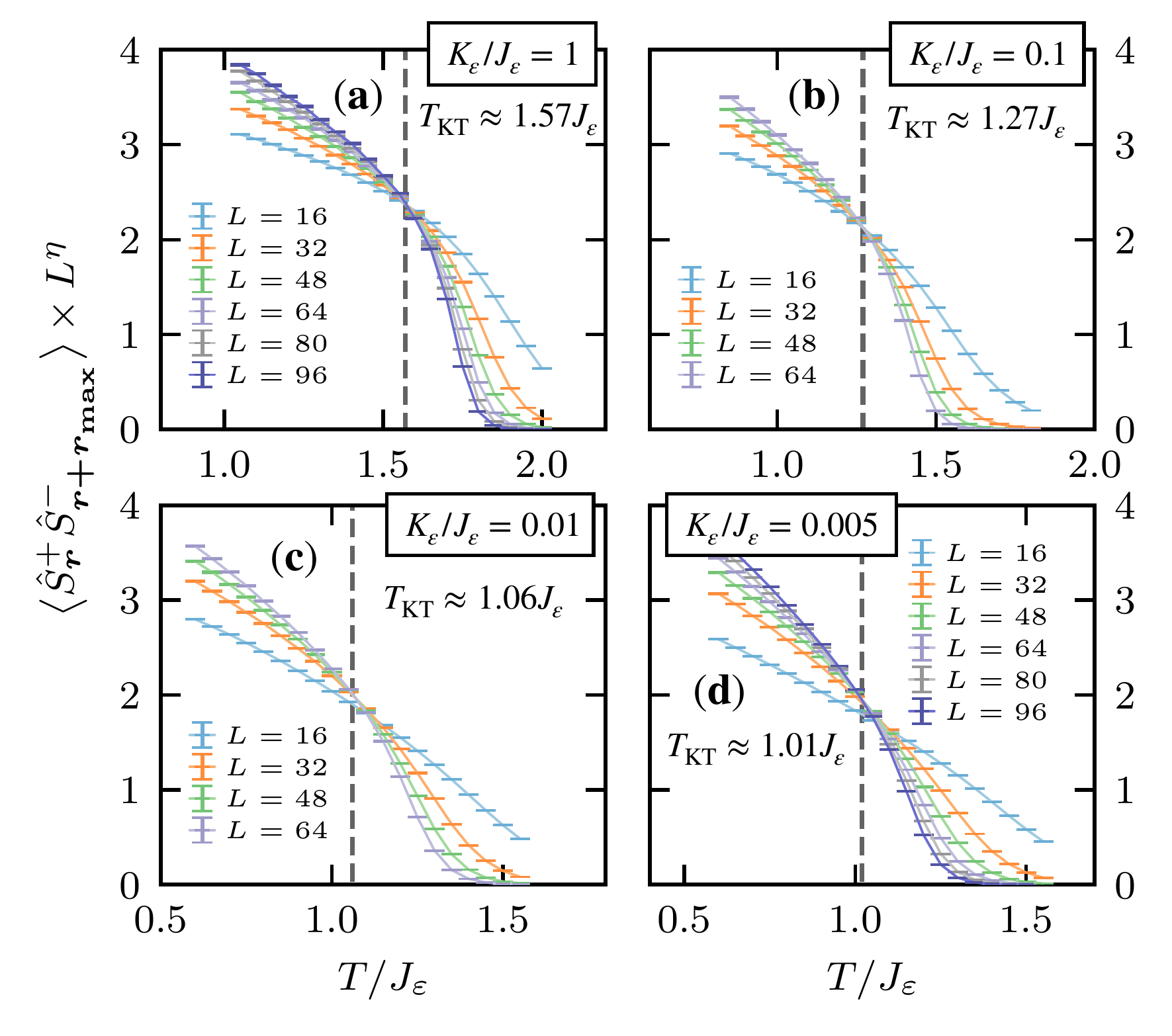} 
    \caption{Two-point off-diagonal spin correlation function at the longest distance $\boldsymbol{r_\mathrm{max}}$ available between two spins (i.e., $\|\boldsymbol{r_\mathrm{max}}\|\propto L$): $\bigl\langle\hat{S}^+_{\boldsymbol{r}}\hat{S}^-_{\boldsymbol{r}+\boldsymbol{r_\mathrm{max}}}\bigr\rangle$ for different values of $K_\varepsilon/J_\varepsilon$ in the BKT regime (they do not correspond to a physical strain $\varepsilon$). At the BKT transition $T\equiv T_\mathrm{BKT}$, one has $\bigl\langle\hat{S}^+_{\boldsymbol{r}}\hat{S}^-_{\boldsymbol{r}+\boldsymbol{r_\mathrm{max}}}\bigr\rangle\sim L^{-\eta}$ with a universal exponent $\eta=1/4$. Here, we use this relation to determine $T_\mathrm{BKT}$. As the ratio $K_\varepsilon/J_\varepsilon$ decreases, systematic drifts of the crossing point with the smallest system sizes are visible.}
    \label{fig:corr_kt}
\end{figure}

The first way we use to determine the BKT transition is based on the finite-size scaling of the spin stiffness $\rho_\mathrm{s}$, as explained in the main text. See Fig.~\ref{fig:nn_bond_4}\,(b,c) of the main text as well as Fig.~\ref{fig:kt_stiffness} for additional data corresponding to strains $\varepsilon=-3.5\%$ and $\varepsilon=-3.0\%$.

Another way to obtain the BKT temperature is by considering the two-point off-diagonal spin correlation function at the longest distance $\boldsymbol{r_\mathrm{max}}$ available between two spins $\bigl\langle\hat{S}^+_{\boldsymbol{r}}\hat{S}^-_{\boldsymbol{r}+\boldsymbol{r_\mathrm{max}}}\bigr\rangle$, with $\|\boldsymbol{r_\mathrm{max}}\|\propto L$. At $T=T_\mathrm{BKT}$, it has the following finite-size scaling,
\begin{equation}
    \bigl\langle\hat{S}^+_{\boldsymbol{r}}\hat{S}^-_{\boldsymbol{r}+\boldsymbol{r_\mathrm{max}}}\bigr\rangle\sim L^{-\eta},
    \label{eq:bkt_corr}
\end{equation}
with $\eta$ the anomalous exponent which takes the value $\eta=1/4$ at the BKT transition. We show results using the relation of Eq.~\eqref{eq:bkt_corr} in Fig.~\ref{fig:corr_kt}. The crossing point of the data for different linear sizes $L$ when rescaling the $y$ axis accordingly signals the BKT temperature $T_\mathrm{BKT}$.

\section{Details of the spin-wave -analysis}
\label{sec:app:swa}

In the FM phase with EA, we apply Holstein-Primakoff transformation~\cite{PhysRev.58.1098} to find the bosonic excitation Hamiltonian. This transformation, up to the first order, is defined as,
\begin{equation}
    \hat{S}^z_{\boldsymbol{r}}= S - \hat{a}^\dag_{\boldsymbol{r}} \hat{a}_{\boldsymbol{r}},\quad\mathrm{and}\quad\hat{S}^-_{\boldsymbol{r}} = \sqrt{2S} \hat{a}^\dag_{\boldsymbol{r}},
    \label{HP}
\end{equation}
where $\hat{a}_{\boldsymbol{r}}$ is a destruction operator of boson at ${\boldsymbol{r}}$ which lies in $\mathcal{A}$ sublattice and the same definition is valid for the other sublattice with $\hat{b}_{\boldsymbol{r'}}$ where $\boldsymbol{r'}$ points out $\mathcal{B}$ sublattice. Collecting all relevant terms and after unitary transformation, we obtain the magnon spectrum of the main text.

However, in the AFM with EP case, we use the MME approach~\cite{Lindgard_1976,Balucani_1980}, as discussed in the main text. Treating the anisotropy as a perturbation and with the small expansion parameter $d_\varepsilon=K_\varepsilon/6J_\varepsilon S$, we find the first-order approximation of the spin-operators in terms of bosonic ones $\hat{a}_{\boldsymbol{r}},\hat{b}_{\boldsymbol{r'}}$,
\begin{equation}\label{MME}
\begin{split}
&
    \hat{S}^z_{\boldsymbol{r}}= S - \hat{a}^\dag_{\boldsymbol{r}} \hat{a}_{\boldsymbol{r}} + d_\varepsilon \sqrt{S\left(S - \frac{1}{2}\right)}\bigl(\hat{a}_{\boldsymbol{r}} \hat{a}_{\boldsymbol{r}} + \hat{a}^\dag_{\boldsymbol{r}} \hat{a}^\dag_{\boldsymbol{r}}\bigr),
    \\&
    \hat{S}^-_{\boldsymbol{r}} = \sqrt{2S} \left[\hat{a}^\dag_{\boldsymbol{r}} - d_\varepsilon\left(S - \frac{1}{2}\right) \, \hat{a}_{\boldsymbol{r}} \right],
    \\&
    \hat{S}^z_{\boldsymbol{r'}}= - S + \hat{b}^\dag_{\boldsymbol{r'}} \hat{b}_{\boldsymbol{r'}} - d_\varepsilon \sqrt{S\left(S - \frac{1}{2}\right)}\bigl(\hat{b}_{\boldsymbol{r'}} \hat{b}_{\boldsymbol{r'}} + \hat{b}^\dag_{\boldsymbol{r'}} \hat{b}^\dag_{\boldsymbol{r'}}\bigr),
    \\&
    \hat{S}^-_{\boldsymbol{r'}}= \sqrt{2S} \left[\hat{b}_{\boldsymbol{r'}} - d_\varepsilon\left(S - \frac{1}{2}\right) \, \hat{b}^\dag_{\boldsymbol{r'}} \right],
    \end{split}
\end{equation}
where ${\boldsymbol{r}}$, ${\boldsymbol{r'}}$ lie in $\mathcal{A}$ and $\mathcal{B}$ sublattices respectively.

Moreover, by expanding the spin-wave dispersion around the $\Gamma$-point, we explore how the dispersion behaves at the bottom of the spectrum. In the FM phase, we find,
\begin{align}
    &
    \omega_\alpha\bigl(\boldsymbol{q}\bigr) = 2 + \frac{K_\varepsilon}{3J_\varepsilon} - \frac{a_{\varepsilon}^2}{4}\boldsymbol{q}^2,
    \\&
    \omega_\beta\bigl(\boldsymbol{q}\bigr) = \frac{K_\varepsilon}{3J_\varepsilon} + \frac{a_{\varepsilon}^2}{4}\boldsymbol{q}^2,
    \end{align}
where $a_{\varepsilon}$ refers to lattice constant. Both bands are quadratic in $\boldsymbol{q}$, see Fig.~4 (a) in the main text. Besides, for AFM phase with EP, up to the first order of $d_\varepsilon$, we obtain,
\begin{align}
    &
    \omega_\alpha\bigl(\boldsymbol{q}\bigr) = 2\sqrt{d_\varepsilon(2S+1)} \left[1-\frac{a_{\varepsilon}^2 \left[d_\varepsilon(2S+1)-1\right]}{16d_\varepsilon\left(2S+1\right)}\boldsymbol{q}^2\right],
    \\&
    \omega_\beta\bigl(\boldsymbol{q}\bigr) = \sqrt{\frac{d_\varepsilon(2S+1)+1}{2}} \, a_{\varepsilon} |\boldsymbol{q}|,
\end{align}
where it shows that the lowest energy $\omega_\beta\bigl(\boldsymbol{q}\bigr)$ is linear in $\boldsymbol{q}$ while $\omega_\alpha\bigl(\boldsymbol{q}\bigr)$ has quadratic dependency, see Fig.~4\,(b) of the main text.

\section{Different exchange paths}
\label{sec:exchange_paths}

\begin{figure}[!h]
    \includegraphics[width=0.8\columnwidth]{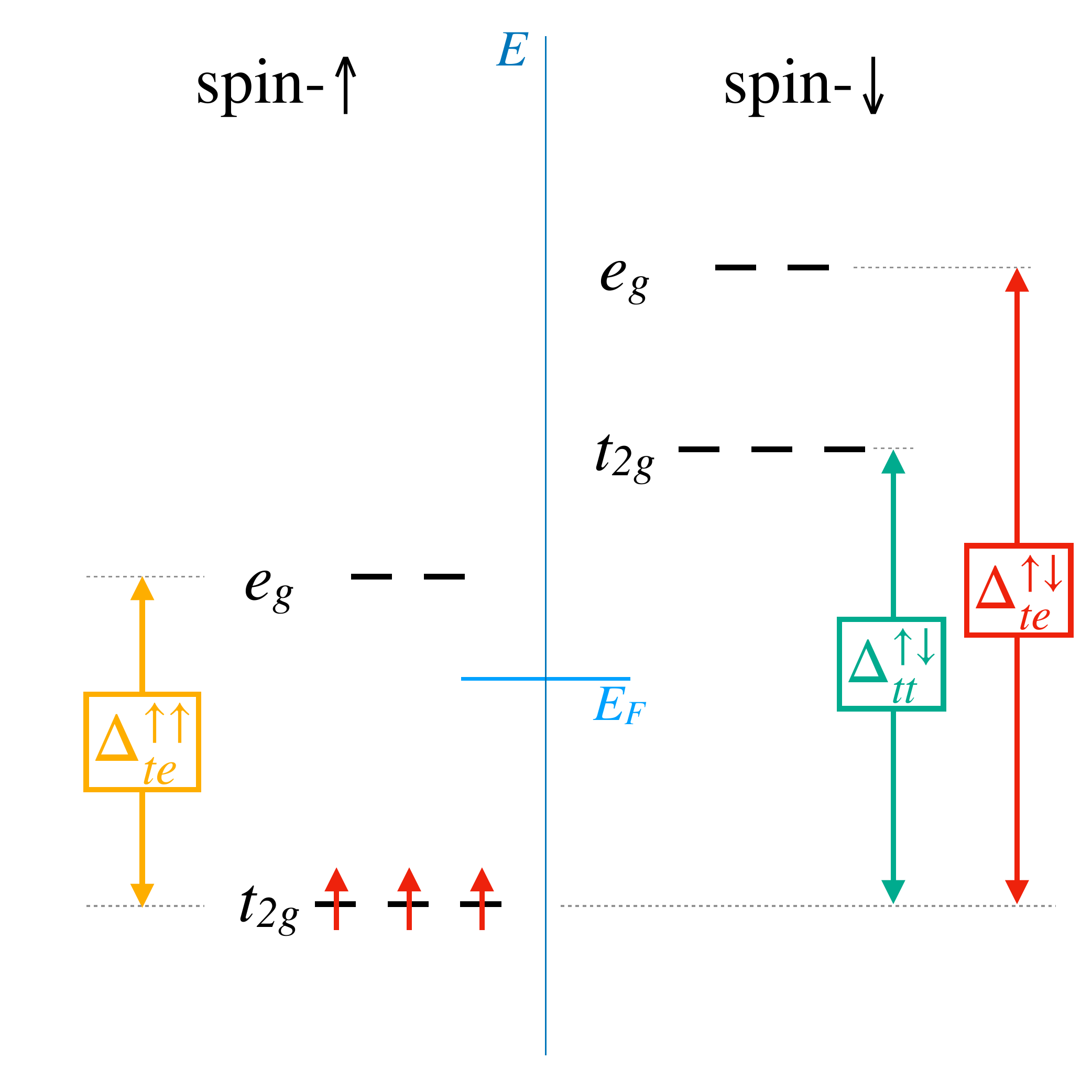}
    \caption{Energy level diagram of Cr-$d$ states in CrCl$_3$.}
    \label{fig:energy_diag}
\end{figure}

Chromium atoms in CrCl$_3$ are surrounded by edge-sharing octahedra, formed by chlorine atoms. The Cr has nominal $3+$ oxidation state, which means that there are three $d$-electrons. A schematic version of the energy level diagram of Cr-$d$ states in CrCl$_3$ calculated in Ref.~\onlinecite{PhysRevB.99.104432} is shown in Fig.~\ref{fig:energy_diag}.

\begin{figure}[!h]
    \includegraphics[width=\columnwidth]{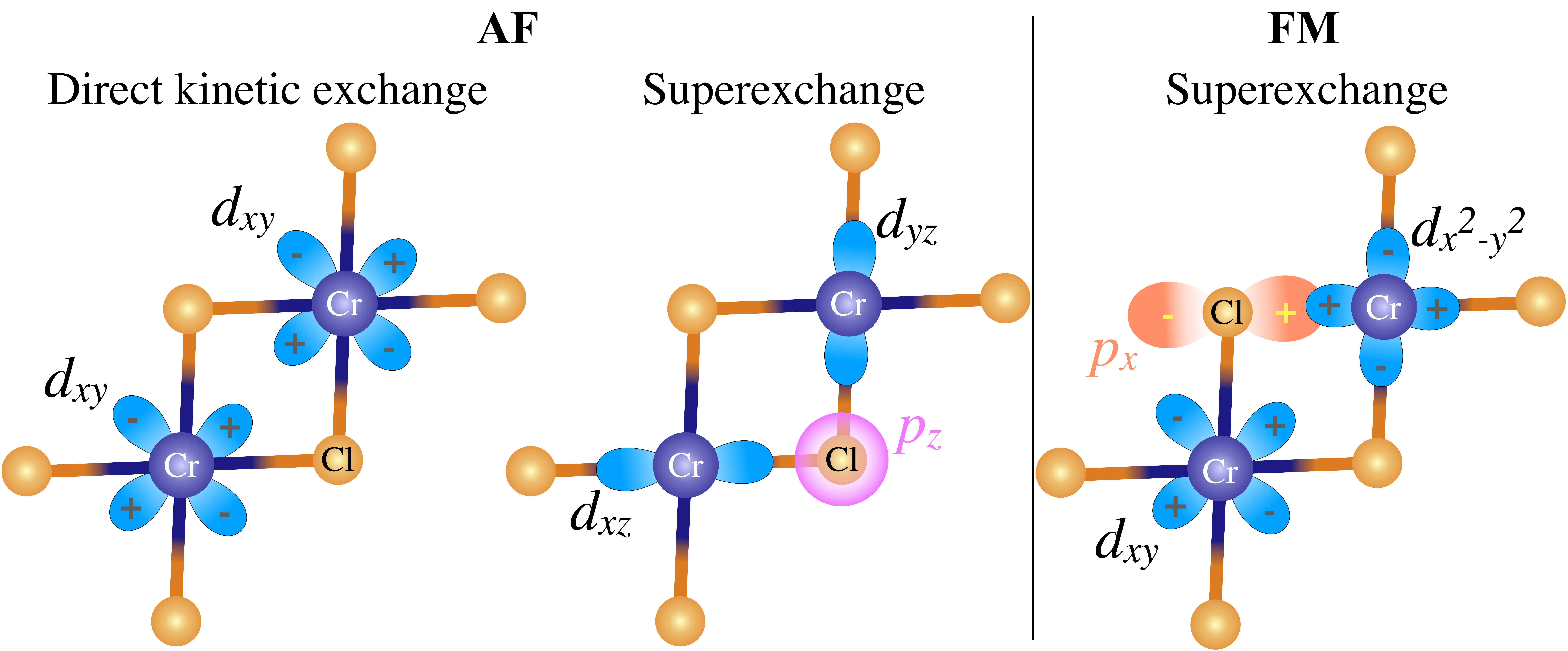}
    \caption{Nearest-neighbour Cr-Cr bond geometry in CrCl$_3$ and the resulting FM and AFM contributions to the exchange coupling.} 
    \label{fig:nn_bond_4}
\end{figure}

The spin-up (majority) and spin-down (minority) states are split by the exchange field, denoted as $\Delta^{\uparrow\downarrow}$. Due to strong crystal field, the electrons primarily occupy the subset of spin-up $t_{2g}$ ($t$) orbitals, forming a half-filled manifold. The $e_g$ ($e$) orbitals are nominally empty and are located right above Fermi level. This splitting between majority-spin $t_{2g}$ and $e_g$ states is thus denoted by $\Delta^{\uparrow\uparrow}_{te}$. In reality the orbitals degeneracy is further lowered due to distortion of octahedra, but we omit this effect for the sake of our argument.
According to the theory of superexchange~\cite{PhysRev.115.2,GOODENOUGH1958287,KANAMORI195987}, one can show that for a nearly 90$^{\circ}$ bond angle the nearest-neighbour exchange interaction will consist of two terms~\cite{PhysRevB.99.104432,Kashin_2020,PhysRevB.102.115162}:
\begin{eqnarray}
    J \propto - \biggl( \frac{1}{\Delta^{\uparrow\uparrow}_{te}} - \frac{1}{\Delta^{\uparrow\downarrow}_{te}} \biggl) t_{te}^2 + \frac{t_{tt}^2}{\Delta^{\uparrow\downarrow}_{tt}},
    \label{eq:nn_j}
\end{eqnarray}
where $t_{te}$ ($t_{tt}$) is the effective hopping integrals between the $t_{2g}$ orbitals on one Cr atom and $e_g$ ($t_{2g}$) states on the neighbouring atom. It is clear that the first term is negative, since $\Delta^{\uparrow\uparrow}_{te} < \Delta^{\uparrow\downarrow}_{te}$, which results in a FM contribution. Thus, virtual hoppings from the $t_{2g}$ electrons to the nominally empty $e_g$ orbitals provide FM superexchange, while the hopping between the $t_{2g}$ states gives rise to an AFM contribution. These contributions to the $J$ coupling are shown in Fig.~\ref{fig:nn_bond_4}.

However, as can be seen in Fig.~\ref{fig:nn_bond_4}, there are two distinct types of hopping processes, the $t_{2g}$ orbitals are involved into. The two $t_{2g}$ orbitals on the neighbouring Cr sites (e.g., $d_{xz}$ and $d_{yz}$) can both hybridize with the same Cl-$3p$ orbital (e.g., $p_z$), which results in indirect hopping processes. At the same time, certain $t_{2g}$ orbitals (e.g., $d_{xy}$ as shown in Fig.~\ref{fig:nn_bond_4}) have their lobes pointing directly towards each other, which gives rise to a ``direct'' kinetic exchange between them.

We interpret our full DFT results in the light of this analytical formula, see Eq.~\eqref{eq:nn_j}. As the material is compressed, the orbital overlap and therefore the hopping amplitudes become larger. The AFM $tt$-derived contribution is therefore expected to grow primarily because of the strong increase of the direct hopping between the orbitals pointing towards each other, which is associated with the kinetic exchange. At the same time, the indirect hopping processes involving Cl-$p$ states, which give rise to FM $te$- and AFM $tt$-derived superexchange are likely to have relatively weaker dependence on strain. Thus, the overall change of sign of $J$ happens because the kinetic exchange between the $t_{2g}$ electrons starts to play the dominant role.

\section{DFT results with PBE functional}
\label{sec:pbe}

We have repeated the DFT calculations employing PBE exchange-correlation functional~\cite{gga-pbe}, which was also used in Ref.~\onlinecite{PhysRevB.98.144411}.

\begin{figure}[!h]
    \includegraphics[width=\columnwidth]{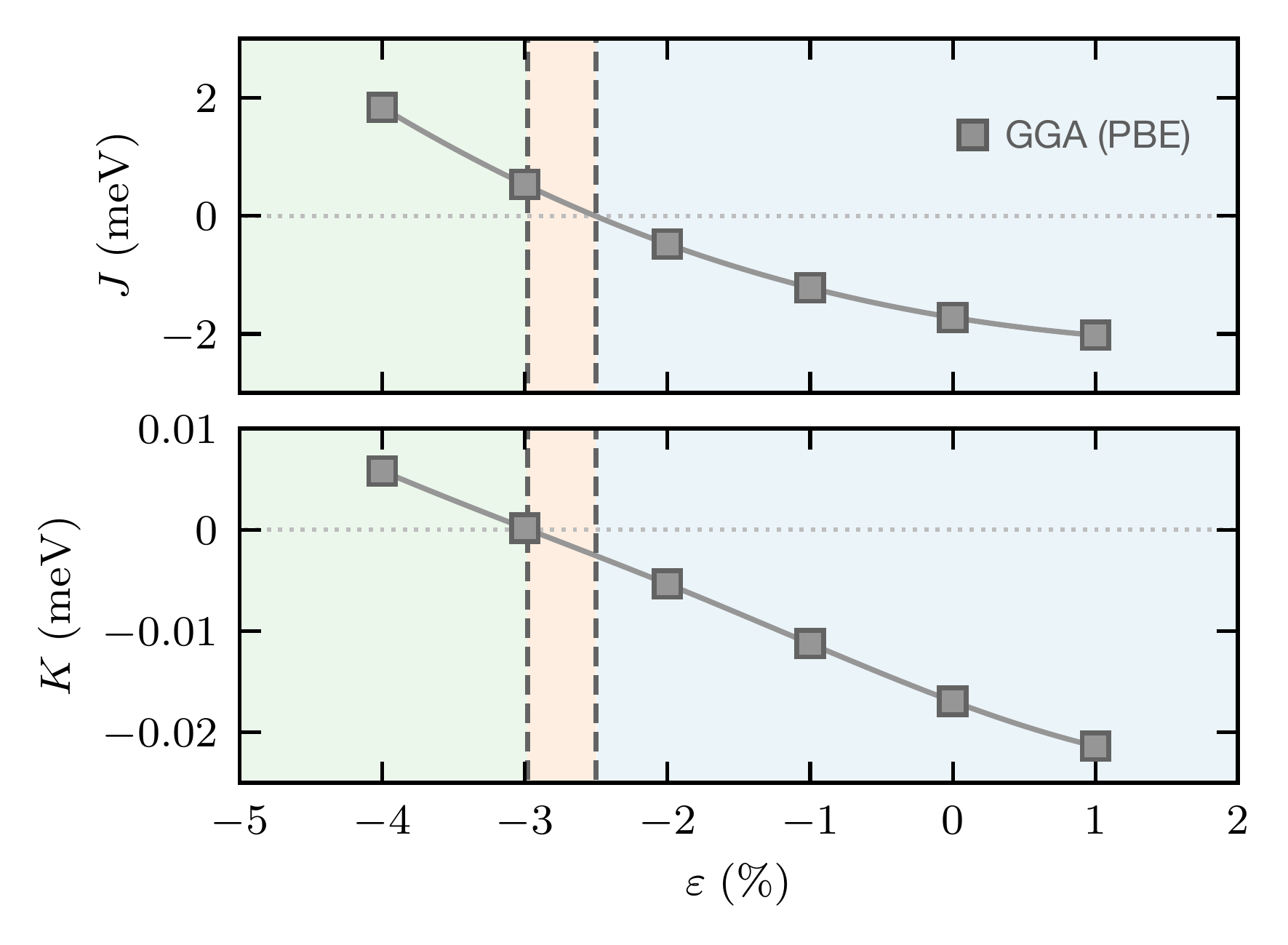}
    \caption{Calculated nearest-neighbour exchange parameter $J_\varepsilon$ and anisotropy constant $K_\varepsilon$ (see Eq.~(1) in the main text for the Hamiltonian definition) as a function of strain $\varepsilon$ in CrCl$_3$ monolayer obtained using PBE functional.} 
    \label{fig:pbexc}
\end{figure}

Inspecting the results shown in Fig.~\ref{fig:pbexc}, one can see that by considering more data points than was done in Ref.~\onlinecite{PhysRevB.98.144411}, we could resolve three phases: BKT, Ising-AFM and Ising-FM ones. Although the Ising-AFM phase is relatively more narrow compared to our prediction from Fig.~1 of the main text, the two functionals result in quantitatively identical phase diagrams.

\end{document}